\documentclass[12pt]{article}

\def\e{\mathrm{e}}
\def\beq{\begin{eqnarray}}
\def\eeq{\end{eqnarray}}

\def\lsim{\mathrel{\rlap{\lower3pt\hbox{\hskip0pt$\sim$}}
     \raise1pt\hbox{$<$}}}         

\usepackage{graphicx}
\usepackage{subfigure}
\usepackage{amsmath}
\usepackage{amsfonts}
\usepackage{amssymb}
\usepackage[T1]{fontenc}
\usepackage{calligra}

\newcommand{\comment}[1]{}

\begin{document}

\begin{titlepage}

\thispagestyle{empty}

\comment{
\begin{flushright}
{NYU-TH-09/04/15}
\end{flushright}
\vskip 0.9cm
}

\topmargin 3cm
\vskip 5cm

\centerline{\Large \bf Two Component Charged Condensate in White Dwarfs}
               
\vskip 0.7cm
\centerline{\large Mehrdad Mirbabayi}
\vskip 0.3cm
\centerline{\em Center for Cosmology and Particle Physics, Department of Physics,}
\centerline{\em New York University, New York, NY  10003, USA}

\vskip 3.cm

\begin{abstract}

The possibility of the formation of a condensate of charged spin-0 nuclei inside white dwarf cores, studied in arXiv:0806.3692 and arXiv:0904.4267, is pursued further. It has been shown, for cores composed mainly of one element (Helium or Carbon), that after condensation phonons become massive and the specific heat drops by about two orders of magnitude. In this note we extend that analysis by considering the coexistence of the nuclei of both types (Helium and Carbon), whose condensation points are generically different. An effective field theory is developed to describe the system when both elements are condensed. The spectrum of fluctuations of this two component charged condensate possesses a collective massless mode with $\omega\propto {\bf k}^2$. Assuming that the fraction of the less abundant element is greater than $\sim 1/100$, the thermal history changes as follows: There is a softer discontinuity in the average specific heat after the condensation of first sector, resulting in slower cooling and a milder drop in luminosity function. The specific heat remains almost constant until the condensation of the second sector, then starts to declines as $T^{3/2}$.

\end{abstract}

\vspace{0cm}

\end{titlepage}

\section{\label{sec:Intro}Introduction and Summary}

In a number of previous papers the possibility of forming a charged condensate in the core of white dwarfs, composed mainly of a single element (Helium or Carbon), and its impact on the cooling process has been studied \cite{Gabadadze:2008mx, Gabadadze:2009dz, Gabadadze:2009jb}. It was observed that upon condensation the phonons which in standard scenarios of white dwarf evolution are the main carriers of internal energy, are eaten by the EM gauge field, become massive and freeze out (see also \cite{Dolgov1}). In this circumstance the relatively small specific heat of the degenerate electrons ($c_v<10^{-2}$) remains as the only contributor to total heat capacity, hence the acceleration of the cooling process.

In this paper we study the condensation of a white dwarf core containing both $He$ and $C$. This can in particular happen in the following situations:

1- Carbon density $\rho_C$ is above $\sim 10^9 gr/cm^3$, so that the critical temperature for condensation, $T_c^C$ \footnotemark, exceeds crystallization temperature $T_{cryst}^C$ \footnotemark, at about $2\times10^7 K$. $\rho_C$ should also remain below the neutronization threshold $\sim 4\times 10^{10}gr/cm^3$. In addition we allow the Helium fraction, $x_{He}\equiv N_{He}/N_{total}$, to range from $\sim 1/100$ (when the heat capacity of the impurity starts to dominate that of electron gas) up to $\sim 1/2$. In this range $T_{cryst}^{He}<T_c^{He}$.

2- When the white dwarf is composed mainly of $He$ at densities higher than or equal to $\sim 10^6gr/cm^3$, as in low mass Helium white dwarfs ($M\lsim 0.5 M_\odot$) \cite{Liebert:2004pj, Strickler:2009kr}. We also assume the existence of Carbon impurities greater than $1/100$. In this case $T_c^C<T_{cryst}^C<T_c^{He}$, so one may expect that after the condensation of $He$ nuclei the $C$ nuclei form a crystal at $T_{cryst}^C$. However due to screening effects of $He$-condensate (studied in \cite{Gabadadze:2008pj,Dolgov2}), the $C$ nuclei barely interact with each other and fail to crystallize\footnotemark.

\footnotetext[1]{
Condensation temperature of a non-interacting Bose gas is given by $T_{c\;(ni)}^X=2\pi n_X^{2/3}/(\zeta(3/2))^{2/3}m_X$, where $m_X$ is the mass and $n_X$ the density of type $X$ ions. However the transition temperature is somewhat higher for interacting particles \cite{Huang:1999zz,Rachel}, and may be approximated as the temperature at which the de Broglie wavelengths of the ions of each type start to overlap: $T_{c\;(i)}^X\simeq 4\pi^2/3m_X d_X^2$, where $d_X=(3/4\pi n_X)^{1/3}$ is the average interparticle separation. This gives about an order of magnitude higher values than $T_{c\;(ni)}^X$.
}

\footnotetext[2]{
The crystallization temperature is the temperature at which the ratio of coulomb interaction of adjacent ions to temperature, $\Gamma$, becomes sufficiently large: $\Gamma \equiv [(Ze)^2/4\pi d]/T_{cryst} \sim 180$.
}

\footnotetext[3]{
Thus, when the second sector forms a small fraction of the the star, it is reasonable to use $T_{c\;(ni)}^X$ instead of $T_{c\;(i)}^X$ as the transition temperature for this sector. To encode this variability we define $\lambda_X$ according to 
\beq
\label{eq:T_c}
T_c^X \equiv\lambda_X\frac{n_X^{2/3}}{m_X}\,.
\eeq
}

Therefore it is likely that the $He$ and $C$ nuclei in the core condense as the temperature drops below their $T_c$. The condensate can well be described by an effective field theory similar to the one developed in \cite{Greiter:1989qb,Gabadadze:2007si}. Nevertheless, after the condensation of both $He$ and $C$, there are two elements involved and in \S2 we will modify the effective theory to describe it. We will show that the two component charged condensate, unlike its one component counterpart, possesses a massless collective mode in the spectrum of fluctuations with $\omega \propto {\bf k}^2$. This mode contributes to specific heat and changes the thermal history significantly.  

The thermal evolution of the white dwarf, analyzed in \S3, would be as follows. At first the system can be approximated by an ideal gas with $c_v=3/2$, and the $\operatorname{log}-\operatorname{log}$ plot of luminosity function (LF) in terms of luminosity has a slope of $-5/7$. Upon the condensation of the first component the corresponding phonons freeze out \cite{Gabadadze:2009dz,Dolgov1}, thus there is a fall in LF curve whose depth depends on the fraction of the second element. The specific heat of the second element's nuclei before condensation is approximately temperature independent ($=3/2$), resulting in a LF curve with the same $-5/7$ slope. Finally as $T_c$ of the second component is approached there is no sharp drop in specific heat because of the existence of the massless mode, rather $c_v$ starts to decrease as $T^{3/2}$, consequently LF curve continuously changes its slope to $-2/7$.

\section{Two Component Charged Condensate}

\subsection{Model and Background}

We use Thomas Fermi approximation to treat electrons as a background density $J_e^0(A_0)$, and associate to each type of ion a complex scalar field with the appropriate charge and mass. In the circumstance of a white dwarf the ions are non-relativistic, therefore the most general effective Lagrangian can be written as \cite{Greiter:1989qb,Gabadadze:2007si}
\beq
\label{eq:lagrangian}
{\cal{L}} =&-&{1\over 4} F_{\mu\nu}^2 
+{\cal P}_1\left(\frac{i}{2}(\Phi^*D_0\Phi-(D_0\Phi)^*\Phi)
               -\frac{|D_j\Phi|^2}{2m_{\Phi}}\right) \nonumber\\
&+&{\cal P}_2\left(\frac{i}{2}(X^*D'_0X-(D'_0X)^*X)
               -\frac{|D'_jX|^2}{2m_{X}}\right)
+ A_{\mu}J_e^{\mu}\,,
\eeq
where $D_{\mu}\equiv \partial_{\mu}-igA_{\mu}$, $D'_{\mu}\equiv \partial_{\mu}-ig'A_{\mu}$, and ${\cal P}_1(x)$ and ${\cal P}_2(x)$ are polynomial functions. 

This system possesses a condensate solution in which the positive charge density of $\Phi$ and $X$ particles is neutralized by the negative charge of electrons, $J_e^0$. To describe the condensate we perform the change of variables: $\Phi \equiv \sqrt{m_{\Phi}}\sigma e^{i\alpha}$, $ X \equiv \sqrt{m_X}\chi e^{i\beta}$, and choose a gauge in which $\alpha=0$, to obtain
\beq 
\label{eq:lagr_cond}
{\cal{L}}&=&-{1\over 4} F_{\mu\nu}^2 
+{\cal{P}}_1\left(gA_0m_\Phi\sigma^2-{1\over2}(\partial_j\sigma)^2
              -{1\over2}g^2\sigma^2A_j^2\right)  \\
&+&{\cal{P}}_2\left(g'\left(A_0-\frac{\partial_0\beta}{g'}\right)m_X\chi^2
              -{1\over2}(\partial_j\chi)^2
              -{1\over2}g'^2\chi^2\left(A_j-\frac{\partial_j\beta}{g'}\right)^2
              \right)
+ A_{\mu}J_e^{\mu}\,, \nonumber
\eeq

Setting all derivatives except $\partial_0 \beta$ equal to zero in the equations of motion, the background solution must satisfy
\beq 
\label{eq:BG}
&A_\mu=\partial_\mu\beta=0\,,&\\
&{\cal P}'_1(0)={\cal P}'_2(0)=1\,,&\\
&J^0_{\Phi}+J^0_X+J^0_e=0\,,&\\
&\text{with}\quad J^0_{\Phi}\equiv gm_{\Phi}\sigma^2\,,
                   \quad J^0_X\equiv g'm_X\chi^2\,.&
\eeq

\subsection{Spectrum}

The spectrum of fluctuations around this background contains two transverse and one longitudinal modes of a massive gauge field, and a Nambu-Goldstone boson. The existence of the latter is easily understood from the symmetries of the problem. The original Lagrangian possesses two global $U(1)$ symmetries corresponding to separate phase rotations of the two complex scalar fields. They, however, are spontaneously broken by the background solution, giving rise to two massless Nambu-Goldstone bosons. Out of these two one combination becomes the longitudinal mode of the massive gauge field, while the other one remains massless. In the limit $m_{\Phi},m_X\to\infty$ the dispersion relations simplify to
\beq
\label{eq:spectrum1}
\omega_T^2&=&{\bf k}^2+m_1^2+m_2^2\,.\\
\omega_L^2&=& {1\over4}\left(\frac{m_1^2}{m_{\Phi}^2}+\frac{m_2^2}{m_X^2}\right)
               \frac{{\bf k}^4}{m_1^2+m_2^2}+m_1^2+m_2^2\,,\\
\label{eq:massless}
\omega_G^2 &=& {1\over4}\left(\frac{m_1^2}{m_X^2}+\frac{m_2^2}{m_\Phi^2}\right)
            \frac{{\bf k}^4}{m_1^2+m_2^2}\equiv \frac{{\bf k}^4}{4M^2}\,,
\eeq
where we have defined
\beq
\label{eq:mdef}
m_1\equiv g\sigma_c\,,~~
m_2\equiv g'\chi_c\,.
\eeq
Had we started from a relativistic Lagrangian, these dispersion relations would have remained unchanged to leading order, except for a ${\bf k}^2$ correction to longitudinal photon. In addition, two heavy modes corresponding to pair productions of $\Phi$ and $X$ particles would have appeared in the spectrum.

Note that when the density of one species tends to zero, $M$ approaches the mass of that species. This confirms that the particles of this type are efficiently screened by the charged condensate of the dominant element, and the massless mode describes the low energy excitations of a non-interacting Bose-Einstein condensate, formed by these screened particles.

\section{White Dwarf Cooling}

Since fusion has terminated inside the white dwarfs, they cool down by radiation according to \cite{Mestel,Shapiro:1983du} 
\beq
\label{eq:cooling}
c_v\frac{dT}{dt}=-\frac{L}{N}=-C (x_{He}A_{He}+x_CA_C)m_uT_*^{7/2}\,,
\eeq
where $c_v$ is the average specific heat per nucleus, $L$ the luminosity, $N$ the total number of nuclei, $A_X$ and $x_X$ the baryon number and fraction of ion $X$, $m_u$ the atomic mass unit, and $C$ a function of atmospheric composition. In this circumstance the luminosity function \cite{Shapiro:1983du}, which is an observable measure of the cooling rate of stars, under the assumption of uniform star formation both in space and time takes the form
\beq
\phi\propto\left[\frac{d\operatorname{log}(L/L_{\odot})}{dt}\right]^{-1}\propto L^{\frac{n(k+1)}{4}-1}\equiv L^{\alpha}\,.
\eeq
Here $k$ and $n$ are defined as exponents describing the temperature dependence of specific heat and luminosity
\beq
c_v \propto T^k\,,~~~~~L\propto T^{4/n}\,.
\eeq
$n$ is $8/7$, thus, $c_v$ which is the single most important variable in white dwarf evolution (given the composition of the star), is read easily from the $\operatorname{log}-\operatorname{log}$ plot of $\phi$ in terms of $L$ (LF curve).

At high temperature it is a good approximation to treat the ions as a classical gas with $c_v=3/2$. This gives the familiar Mestel cooling curve with $\alpha=-5/7$. As the temperature drops below the $T_c^X=\operatorname{max}(T_c^{He},T_c^C)$ (see footnote 1) the $X$ nuclei cease to contribute to specific heat because of the gap emerging in the phonon dispersion relation (as explained in \cite{Gabadadze:2009dz}). Nevertheless the specific heat of the other species ($Y$) which is still in the gas state ($c_v^Y=3/2$) amounts effectively to $c_v=3x_Y/2$ per ion. Therefore the LF curve drops by $\operatorname{log}(x_Y)$ units. The curve continues with the same slope $\alpha=-5/7$ afterwards. 

As $T_c^Y$ is approached the model of two component charged condensate can be applied to the system. Except for the massless mode \eqref{eq:massless} in the spectrum of fluctuations, all the other modes have gaps larger than the temperature and are not excited significantly. Therefore the specific heat per ion can be written as
\beq
\label{eq:c_v}
c_v= \frac{\partial}{\partial T} \left[
    \frac{1}{n}\int_0^{k_{max}}\frac{d^3{\bf k}}{(2\pi)^3}
    \frac{\omega_G}{\e^{\omega_G/T}-1}\right]\,,
\eeq
where $n\equiv N/V$ is the total number density of ions (including both $He$ and $C$ nuclei), and $k_{max}$ is of order of the inverse interparticle separation of the less abundant element, above which the effective field theory description breaks down.

\begin{figure}[h!]
	\begin{center}
{\includegraphics[width=10cm,height=7cm]{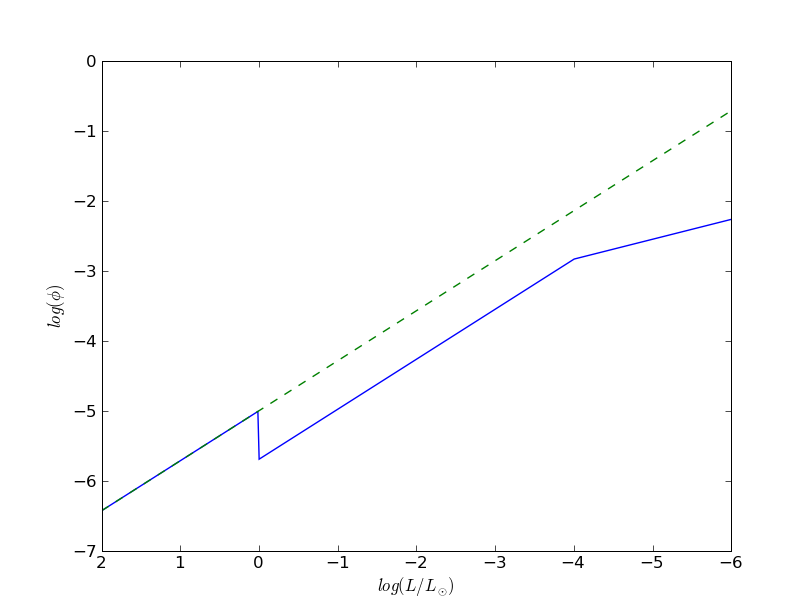}}
        \end{center}
	\caption{\small{Schematic behavior of luminosity function. 
                  The dashed line corresponds to Mestel regime while the solid 
                  line exhibits the condensate behavior far from the 
                  transition point.}}
	\label{fig:luminosity}
\end{figure}

Well below $T_c^Y$ the upper limit of integral can be taken to infinity yielding
\beq
\label{eq:c_v_lowT}
c_v=\frac{15\zeta(5/2)}{32 n} \left(\frac{2MT}{\pi}\right)^{3/2}
  = 5\pi\zeta(5/2)x_Y\left(\frac{2\pi MT}{m_YT_c^Y}\right)^{3/2}\,,
\eeq
and since $m_{He}<M<m_C$, there is no dramatic change in specific heat right after second condensation. Nevertheless it starts to decline as $T^{3/2}$, therefore the LF curve after about
\beq
\label{eq:TX-TY}
\operatorname{log}\left(\frac{L(T_c^X)}{L(T_c^Y)}\right) 
  = \frac{7}{3}\operatorname{log}\left(\frac{n_X}{n_Y}\right)
     -\frac{7}{2}\operatorname{log}\left(\frac{m_X}{m_Y}\right)
     +\frac{7}{2}\operatorname{log}\left(\frac{\lambda_X}{\lambda_Y}\right)\,,
\eeq
(see Eq. \eqref{eq:T_c}) from the first condensation changes slope to $\alpha=-2/7$. 

We conclude by giving two concrete examples and a schematic diagram of luminosity fuction Fig. \ref{fig:luminosity} (note that the typical variations of specific heat at transition points may significantly change the behavior near those points).

As the first example consider a Carbon dominated superdense white dwarf with $\rho \sim 10^{10} gr/cm^3$, $x_{He}\simeq 0.1$, and an envelope composition of
\beq
X\simeq 0\,,\qquad Y\simeq 0.9\,,\qquad Z_m\simeq 0.1\,,
\eeq
where $X, Y$ and $Z_m$ are $H$ fraction, $He$ fraction and metallicity, respectively. Carbon nuclei first reach their condensation temperature at $L\simeq 3 L_\odot$ when LF curve drops for $\operatorname{log}(1/x_{He})=1$ unit, and the second condensation takes place later at $L\simeq 3\times 10^{-4} L_\odot$.

Next consider a Helium dominated white dwarf with $\rho \sim 10^6 gr/cm^3$, $x_C\sim 0.1$, and atmospheric composition of
\beq
X\simeq 0.99\,,\qquad Y\simeq 0\,,\qquad Z_m\simeq 2\times 10^{-4}\,.
\eeq
In this case $He$ condenses first at $L\simeq 1.5\times 10^{-4}L_\odot$ with again a unit drop in LF curve, and then $C$ nuclei condense when $L\simeq 6\times10^{-12}L_\odot$.

\subsection*{Acknowledgements}

I am indebted to Gregory Gabadadze for his guidance, and Rachel Rosen for detailed comments on the manuscript. MM was partially supported by the NSF (grant AST-0908357), NASA (grant NNX08AJ48G), MacCracken Fellowship and James Arthur Graduate Award at NYU.

\end{document}